\documentclass[aps,prl,reprint,superscriptaddress]{revtex4-1}

\usepackage[T1]{fontenc}        
\usepackage{geometry} 
\usepackage{graphicx}
\usepackage[english]{babel}   
\usepackage{amssymb}
\usepackage{epstopdf}
\usepackage{tikz}
\usetikzlibrary{snakes,arrows,shapes,backgrounds}
\usepackage{pgfplots}
\usetikzlibrary{plotmarks,calc,trees,positioning,arrows,chains,shapes.geometric,%
    decorations.pathreplacing,decorations.pathmorphing,shapes,%
    matrix,shapes.symbols}
\usepackage[applemac]{inputenc}
\usepackage{amsmath,amssymb,bm,times}
\usepackage{subfigure}
\usepackage{xcolor}
\usepackage{fullpage}

\graphicspath{{figures/}}

\begin{document}


\title{
Nonlinear Hysteretic Torsional Waves
}


\author{J. Cabaret, P. B\'equin, G. Theocharis}
\affiliation{LUNAM Universit\'es, CNRS, Universit\'e du Maine, LAUM UMR-CNRS 6613, Av. O. Messiaen, 72085 Le Mans, France.}

\author{V. Andreev}
\affiliation{Acoustics Department, Faculty of Physics, Moscow State University, Moscow, Russia}

\author{V.E. Gusev}
\affiliation{LUNAM Universit\'es, CNRS, Universit\'e du Maine, LAUM UMR-CNRS 6613, Av. O. Messiaen, 72085 Le Mans, France.}

\author{V. Tournat}
\email[Corresponding author V. Tournat: ]{vincent.tournat@univ-lemans.fr}
\affiliation{LUNAM Universit\'es, CNRS, Universit\'e du Maine, LAUM UMR-CNRS 6613, Av. O. Messiaen, 72085 Le Mans, France.}


\date{\today}

\begin{abstract}
We theoretically study and experimentally report the propagation of nonlinear hysteretic torsional pulses in a vertical granular chain made of cm-scale, self-hanged magnetic beads. As predicted by contact mechanics, the torsional coupling between two beads is found nonlinear hysteretic. This results in a nonlinear pulse distortion essentially different from the distortion predicted by classical nonlinearities, and in a complex dynamic response depending on the history of the wave particle angular velocity. Both are consistent with the predictions of purely hysteretic nonlinear elasticity and the Preisach-Mayergoyz hysteresis model, providing the opportunity to study the phenomenon of nonlinear dynamic hysteresis in the absence of other type of material nonlinearities. The proposed configuration reveals a plethora of interesting phenomena including giant amplitude-dependent attenuation, short term memory as well as dispersive properties. Thus, it could find interesting applications in nonlinear wave control devices such as strong amplitude-dependent filters.
\end{abstract}


\maketitle

Nonlinear dynamic hysteresis is involved in a wide range of acoustic effects recently observed in complex solids, often called mesoscopic solids \cite{livreguyer}. It is shown to originate from clapping and friction phenomena, induced by acoustic waves at the internal contacts and inter-grain boundaries of polycrystalline and Earth materials \cite{nazarov,nazarov2,johnson,guyerprl99}, damaged solids \cite{can,vdabeele,moussatov} and "model" granular media \cite{inserra,JiaJohnson}. This hysteretic nonlinearity enriches the phenomenology of the classical nonlinearity (small expansion terms of the smooth nonlinear stress-strain relation and geometric nonlinearity) with phenomena like nonlinear attenuation \cite{nazarov,nazarov2}, memory \cite{mccall} and slow dynamics \cite{tencate,johnsonsutin}.

One of the first observed manifestations of mechanical hysteresis is the shear coupling between two elastic spheres in contact \cite{kljohnson,exp}. Interestingly, periodic line-assemblies of spheres, also called granular chains, have attracted a growing interest in the last years for their rich nonlinear dynamics phenomena (harmonic generation, solitons, breathers...)\cite{nester,cabaret,boechler,sen,job} and the potentially wide applications in wave control devices \cite{georgios}. However, up to now, dynamic hysteresis has not been studied in granular chains. In addition, all the media exhibiting dynamic hysteresis (rocks, disordered granular media, concrete, composites...) have been also exhibiting at the same time other types of nonlinearities, classical or nonclassical. Consequently, although some modeling has been proposed \cite{gusev1,gusev2}, it has not been possible yet to observe pulse wave distortion by purely hysteretic nonlinearity. Manifestations of hysteretic nonlinearity were observed only in narrow frequency band experiments with sine waves, via for instance the shift of vibration resonances \cite{nazarov,johnsonsutin} but never in the distortion of pulsed acoustic signals. The transformation of pulse profiles in media with classical quadratic nonlinearity, leading to weak shock front formation of the particle velocity profile, is one of the most classical observations in nonlinear acoustics \cite{mendousse,livrehamilton}. 
Interestingly, the pulse distortion in media with cubic elastic nonlinearity, was observed for the first time only recently \cite{catheline}. 

In this work, we report for the first time the transformation of pulse profile in a medium with pure hysteretic quadratic nonlinearity, essentially different from the distortion by classical nonlinearities.
\begin{figure}
\centering
\includegraphics[width=7cm]{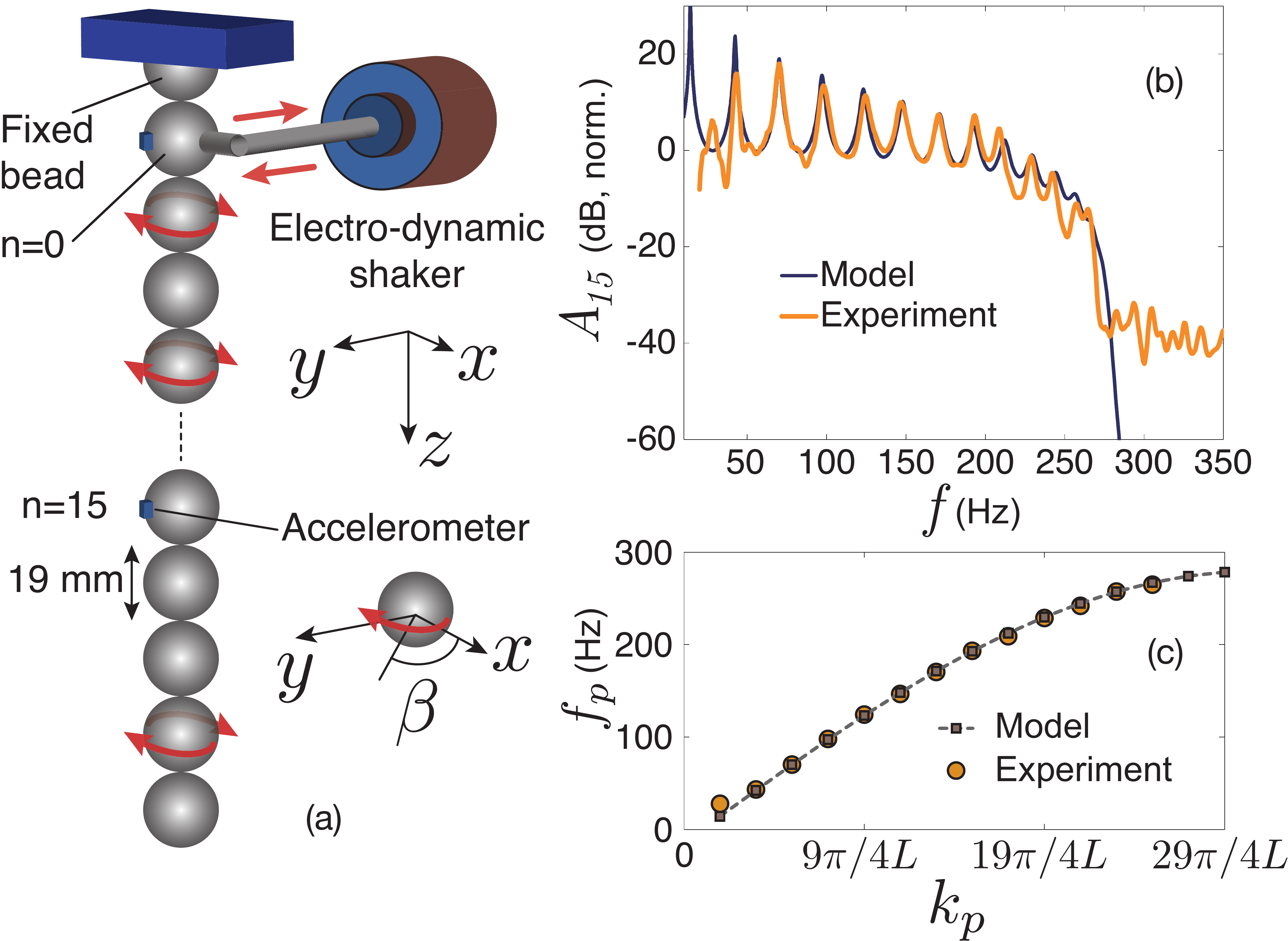}
\caption{(a) Experimental configuration for the torsional wave propagation (pure rotation of the beads around the axis $z$ of the chain). (b) Signal acceleration at bead 15 versus frequency for a 15-bead chain. (c) Dispersion relation retrieved from the $p$ resonance frequencies of the chain. \label{fig1}}
\end{figure}
We start with the observation and characterization of torsional (or pure rotational) wave propagation in a granular chain composed of identical magnetic spheres. Due to the pure torsional coupling at the contacts excited in a rotational motion along the axis $z$ of the chain, pure nonlinear hysteretic behavior is observed. The latter is quantitatively characterized for a single contact in a resonance experiment, and then used for the modeling of torsional wave pulse distortion by quadratic hysteresis. A quantitative comparison with the experimental distorted pulse profiles in a chain of 70 beads, shows the validity of the developed modeling. Extremely large nonlinear self-action (including self-attenuation and pulse deceleration) is demonstrated for parts of the pulse profile, depending on the previous loading by the pulse itself (a short term memory).

{\it Experimental configuration. }The medium is composed of 15 to 70 magnetic beads of diameter $d=2R=19$ mm and mass $m=27$ g (see Fig.\ref{fig1}(a)) \cite{refmagnet,sierra}. The magnetic poles of the beads are aligned along the chain ($z$-axis) and provide a magnetic force $F_0= 54$ N between beads, normal to the contacts. $F_0$ exceeds largely the gravity force and thus the contact forces along the chain are considered identical in this study. The chain is excited at the top by a shaker (B\&K 4810) coupled to the side of the bead $n=0$, which produces a rotational motion relative to the $z$-axis of the chain. This dynamic rotation is transmitted along the chain thanks to the torsional rigidity of the contacts, leading to torsional wave propagation. For detecting the waves, accelerometers (type PCB 352C23) are glued to the side of several beads and angular particle accelerations are recorded via an oscilloscope or a spectrum analyzer. 

The dynamics of the chain can be described through the following system of equations of rotational motion for each bead $n$ interacting with its nearest neighbors $n-1$ and $n+1$ through torsional moments, 
\begin{equation}
J\frac{\partial^2 \beta_n}{\partial t^2} = M_n - M_{n-1} ,
\end{equation}
where $\beta_n$ is the rotation angle of bead $n$, $J=2mR^2/5\simeq 9.85.10^{-7}\ $kg.m$^2$ is the moment of inertia of the beads, $t$ the time, and $M_n$ the nonlinear torsional moment depending on the relative rotation angle between beads $n+1$ and $n$. Following the approach in \cite{deresiewicz,kljohnson}, the moment-angle relationship for oscillatory motion (a simple periodic driving) between two spheres in contact can be approximated by a linear and a quadratic hysteretic functions (see the Supplementary Material),
\begin{eqnarray}\label{eq2}
M_n\!&=&M_n^{lin}+M_n^h \\
&=&\!K_t \! \left\{\! \psi_n\!-\! h\!\!\left[\psi_n^*\psi_n\!+\!\frac{1}{2}\!\left( \psi_n^2-\psi_n^{*2}\right) \! sign(\dot{\psi}_n)\right]\! \right\}\!, \nonumber
\end{eqnarray}
where $\psi_n= \beta_{n+1}-\beta_{n}$ is the relative rotation angle between two adjacent beads and $\psi_n^*$ is its magnitude, $K_t=d(1-\nu) F_0$ is the linear torsional constant, $h=Ea^2/(1+\nu) \mu F_0$ is defined as the parameter of quadratic hysteretic nonlinearity, with $\mu$ the a priori unknown friction coefficient of the bead's material, $E=160$ GPa the Young's modulus, $\nu \simeq 0.24$ the Poisson ratio of the bead's material, and $a=(3F_0R(1-\nu^2)/4E)^{1/3}\simeq 130\ \mu$m  the contact radius.

In Fig.~\ref{fig1}(b), the angular particle acceleration $A_{15}$ of bead $n=$ 15, is presented versus frequency in the case of a 15 bead chain. In this finite-length chain, resonances are observed up to the cut-off frequency $f_c\simeq 275$ Hz. Simulating the linear dynamics of the chain with springs having a complex constant $K_t=0.78 (1+0.027i)$ N.m to fit the experimental losses, we obtain the theoretical curve in Fig.~\ref{fig1}(b). The theoretical cut-off frequency $f_c=\sqrt{K_t/J}/\pi=283\ $Hz is found in good agreement with the experimental one. From the experimental resonance frequencies $f_p$ (with $p=1, \dots, 15$) of the finite chain, the dispersion relation can be retrieved. At the $p$-order resonance of the chain, the wavenumber is $k_p=(2p-1)\pi /4L$, where $L$ is the length of the chain. The relation $f_p-k_p$ is evenly derived for the simulation results and the two dispersion curves are compared successfully in Fig.~\ref{fig1}(c). These observations provide an estimate of the long-wavelength torsional wave velocity in such system $c=d\sqrt{K_t/J}\simeq 17\ m/s$.
\begin{figure}[h!]
\centering
\includegraphics[width=7cm]{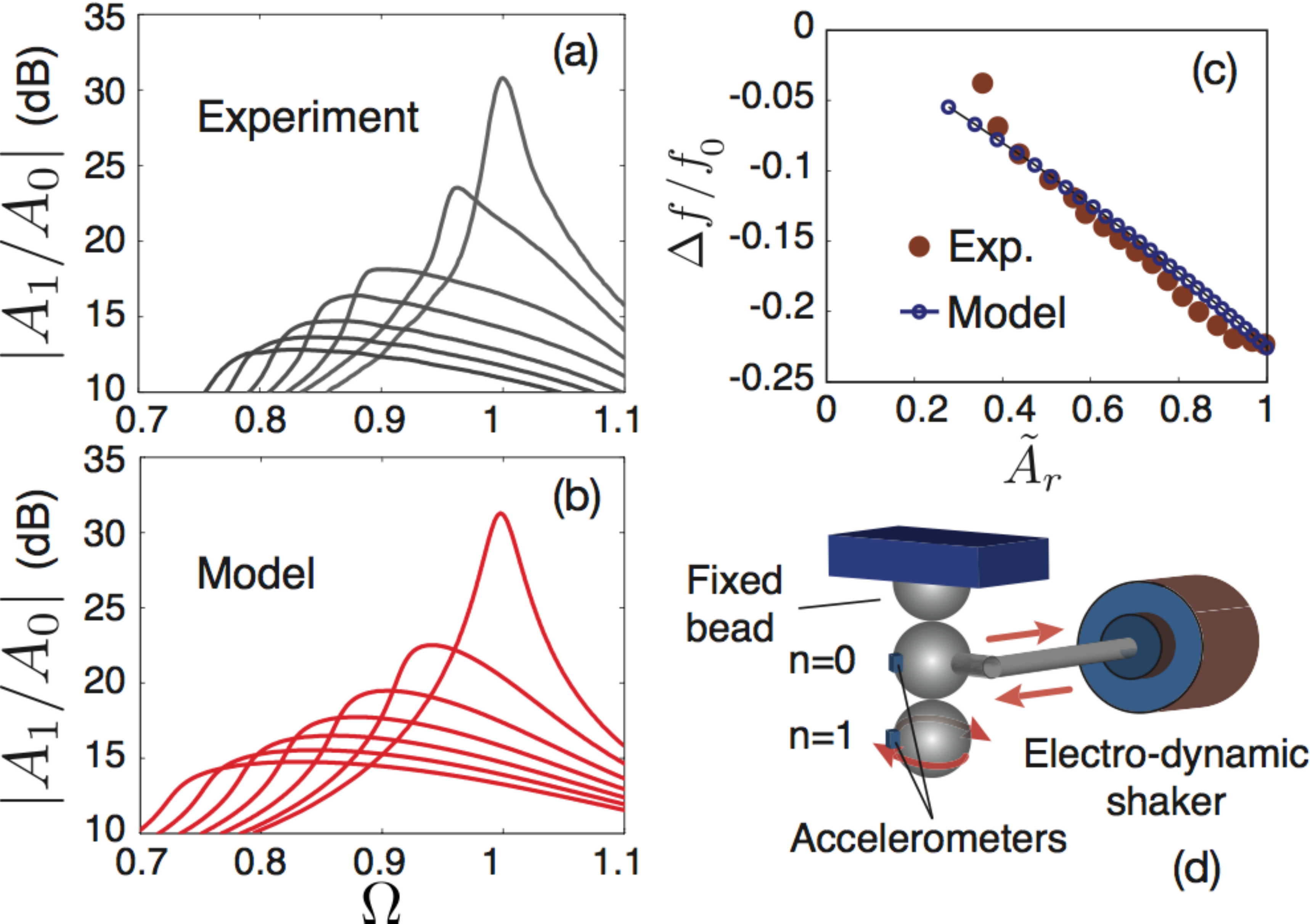}
\caption{Magnitude of the acceleration transfer function (detected acceleration $A_1$ at bead $1$ over the acceleration $A_0$ at bead $0$) of a "2 beads-1 contact" system for different excitation levels, experiment (a) and theory (b). (c) Relative resonance frequency shift as a function of the detected resonance amplitude (normalized to the one at the maximum excitation level). (d) Setup for the single contact characterization in torsion. \label{fig2}}
\end{figure}

In order to verify in our configuration the validity of the nonlinear moment-angle relationship (\ref{eq2}), and to estimate the parameter $h$, we characterize a single contact between two spheres in a nonlinear resonance experiment for rotational motion as depicted in Fig.~\ref{fig2}(d). The nonlinear resonance method has been widely implemented for characterizing the hysteretic behavior because the classical quadratic nonlinearity does not contribute to the amplitude dependent resonance shift at the leading order \cite{johnsonsutin,inserra}. 
The downshift of the resonance frequency with a linear dependence on the oscillation amplitude can be attributed to hysteretic quadratic nonlinearity \cite{nazarov,nazarov2,johnson,guyerprl99}, while the shift proportional to the square of the amplitude at lowest excitation levels - either to cubic elastic nonlinearity \cite{nowick,nazarov4,tencate2} or to hysteretic nonlinearity described by Preisach-Arrhenius model \cite{gusev5}. 

In Fig.~\ref{fig2}(a), the experimental ratio between acceleration signals at bead $0$ and bead $1$ is shown as a function of normalized frequency $\Omega = f / f_0$, where $f_0=159$ Hz is the experimental resonance frequency at the smallest excitation amplitude. Downward resonance frequency shift is observed as well as a nonlinear attenuation process \cite{nazarov,nazarov2}. The relative resonance frequency shift $\Delta f / f_0= (f_r-f_0)/f_0$ is shown to scale linearly with the detected resonance amplitude in Fig.~\ref{fig2}(c), which is consistent with quadratic hysteresis. For comparison, the theoretical transfer function obtained with the harmonic balance method for the moment-angle relationship Eq.~(\ref{eq2}), and neglecting contributions from higher harmonics \cite{NLvibs,gusev6}, is plotted in Fig.~\ref{fig2}(b). A quantitative agreement between the theoretical and experimental resonance curves and resonance frequency shifts is obtained for $h\simeq 258 \pm 16 $ (which corresponds to a realistic friction coefficient $\mu \simeq 0.32 \pm 0.02$). The fitted quality factor $Q_0 = 37$ is also in agreement with the imaginary part of $K_t$ used above.
\begin{figure}[!ht]
\centering
\includegraphics[width=6cm]{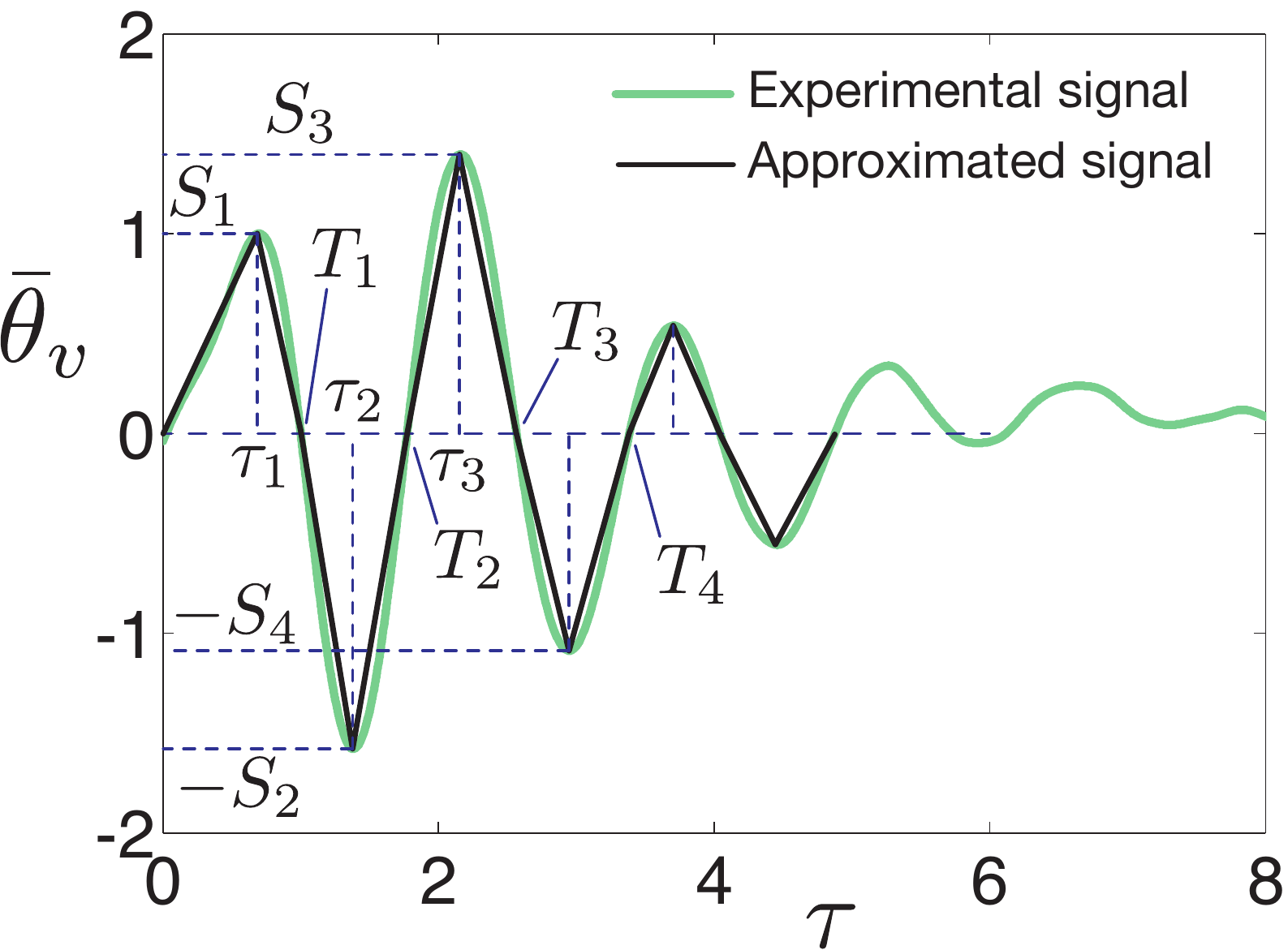}
\caption{Experimental angular particle velocity $\bar{\theta}_v=\theta_v/\theta_0$ at bead 1 for the smallest excitation amplitude and its approximation by a linear piece-wise function. $T_n$ are the positions of the signal zeros in time normalized by $t_0=5.77$ ms so that $T_1=1$, $\tau_n$ are the extrema positions in time, and $S_n$ are the extrema amplitudes normalized by $\theta_0$ so that $S_1=1$.\label{fig3}}
\end{figure}

After having extracted the linear properties of the medium and tested the quadratic hysteretic behavior for a single contact, we turn to the nonlinear propagation of torsional wave pulses. In order to avoid simultaneous detection of forward and backward propagating waves, and to mimic a semi-infinite chain, we use a 70-bead long chain. The detector is placed at the bead $n=$ 18, and the wave propagates through 104 beads before coming back to the receiver after reflection by the free boundary. The pulse central frequency should be sufficiently low to avoid strong dispersion effects occurring at frequencies close to the Bragg frequency $f_c\simeq 275$ Hz, but also sufficiently high to limit its spatial extent and be able to distinguish incident from reflected pulses. Therefore a central frequency of 100 Hz has been chosen. The first bead velocity is shown in Fig.~\ref{fig3}, corresponding to the time-integrated accelerometer signal.

We describe the pulse distortion starting from the following evolution equation, valid for any type of signal in a one-dimensional dispersionless medium where linear dissipation is neglected \cite{gusev1,gusev2,gusev6},
\begin{equation}\label{eq3}
\frac{\partial \bar{\theta}_v}{\partial \xi} - \frac{1}{2}\frac{\partial \bar{M}_h}{\partial \bar{\theta}_v} \frac{\partial \bar{\theta}_v}{\partial \tau} = 0 .
\end{equation}
Here, in accordance with the considered problem, $\bar{\theta}_v=\theta_v/\theta_0$ is the normalized angular particle velocity, $\theta_0$ is the angular velocity amplitude of the first phase of the emitted wave packet at $z=0$, $\tau = (t-z/c)/t_0$ where $t_0=5.77$ ms is the duration of the first phase of the emitted signal, $\bar{M}_h=M_h/(dK_t)$ and $\xi=z/z_{nl}$ with $z_{nl}= c^2 t_0/ hd\theta_0$ the characteristic nonlinear length \cite{note}.

The important term $\partial \bar{M}_h/\partial \bar{\theta}_v$, having the physical sense of a normalized amplitude-dependent modulus \cite{gusev1,gusev2,aleshin}, and containing the hysteretic behavior, has been analytically derived from the Preisach-Mayergoyz (PM) model of hysteresis \cite{livreguyer,preisach,mayergoyz}. This phenomenological model, developed initially in magnetism and then adapted to elastic waves, considers a large number of hysteretic elements (hysterons), each having two possible stress states. The transitions from one state to another take place at two characteristic strains, one for each sign of the strain rate. An hysteretic medium can be represented via a distribution of hysterons in the PM plane formed by two axes whose coordinates are the two characteristic strains of the hysterons. Unlike existing analytical formulas for the quadratic hysteresis, this model has the ability to model the instantaneous memory stored in the hysterons' states under the action of an arbitrary varying acoustic loading and thus, is applicable to arbitrary signals such as the pulses observed here.
 
The evolution equation (\ref{eq3}) is then modified into a system of equations, derived for the linear piece-wise approximated signal of Fig.~\ref{fig3}. The derivation using the PM model of hysteresis is detailed in the Supplementary Materials, and leads to the following system,
\begin{eqnarray}
dT_n&=&S_nd\xi /2 ,\label{eq4}\\
d\tau_n&=&(T_n-\tau_n)A_n d\xi / (T_n-T_{n-1}) ,\label{eq5}\\
dS_n&=& -A_nS_nd\xi /(T_n-T_{n-1}) , \label{eq6}
\end{eqnarray}
with $A_n=S_n$ if $S_n>S_{n-1}$ and $A_n=(S_n+S_{n-1})/2$ if $S_n<S_{n-1}$. The system (\ref{eq4})-(\ref{eq6}) describes respectively the small shifts of the positions in time of the zeros ($dT_n$), the extrema of the signal ($d\tau_n$), as well as the change in extremal values ($dS_n$) when a small change in $d\xi$ occurs, i.e. either a small distance or a small excitation amplitude change. So, by increasing step by step the amplitude $\xi$, the system (\ref{eq4})-(\ref{eq6}) describes the distortion by the hysteretic nonlinearity of the linear piece-wise approximated pulse. We note here that the evolution of the parameters of the signal phase number $n$ depends in general on phases $n$ and $n-1$ but not on previous ($n-2$, $n-3$...). Equations (\ref{eq4}) and (\ref{eq5}) have a clear physical sense. The local time delays are proportional to wave amplitudes as it could be expected for quadratic hysteretic nonlinearity \cite{livreguyer}. Equation (\ref{eq6}) describes local nonlinear absorption with an absorption coefficient proportional to wave amplitude and to the local "frequency" (inverse duration of the pulse phase), also theoretically expected for quadratic hysteretic nonlinearity \cite{johnsonsutin}. Interesting compression/stretching effects could take place for slowly modulated wave packets containing many carrier periods.

\begin{figure}[!ht]
\centering
\includegraphics[width=7cm]{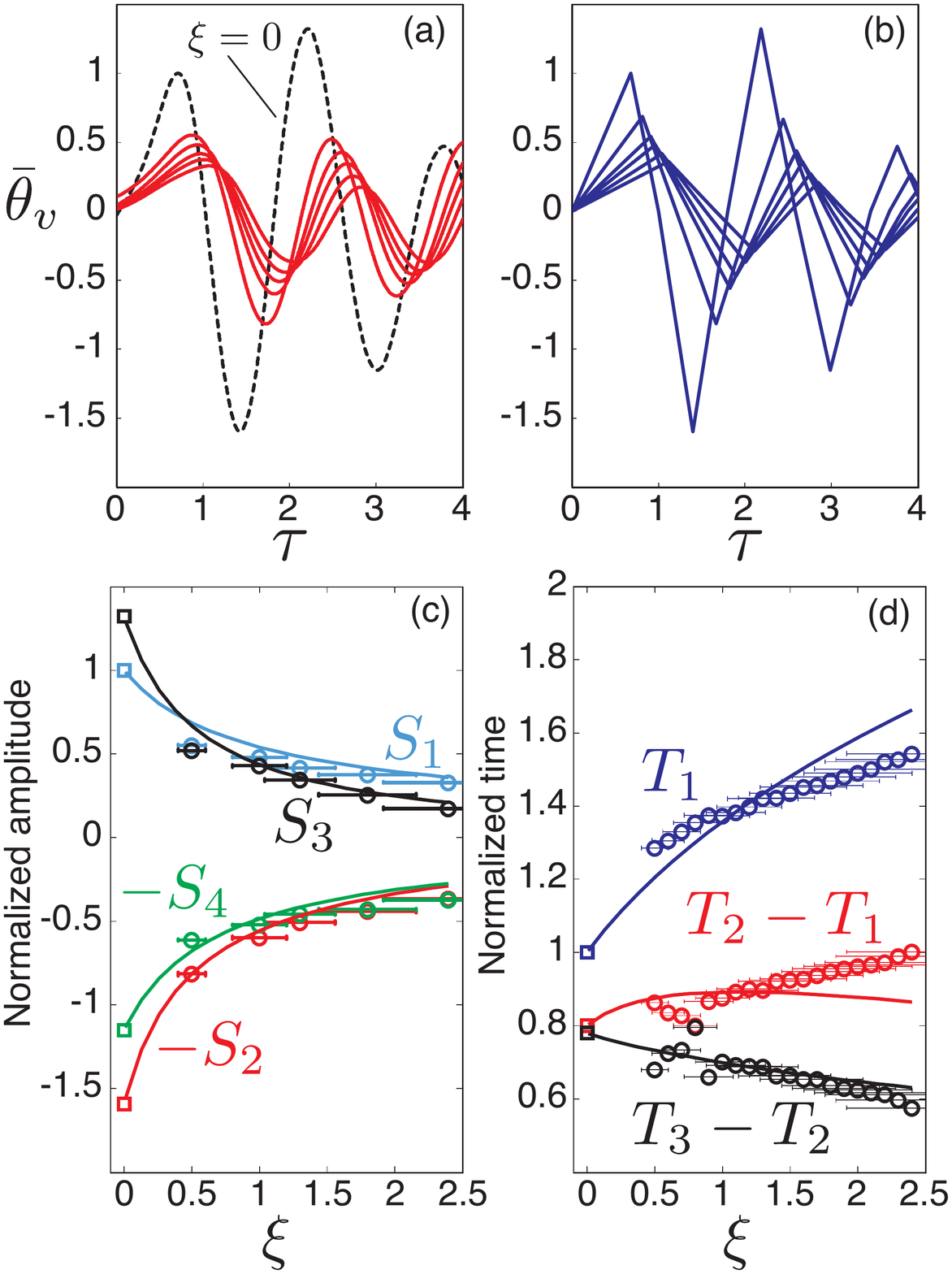}
\caption{(a) Experimental angular velocity signals for different values of $\xi$ : at bead 1 ($\xi=0$ dashed line) and at bead 18 for different excitation amplitudes ($\xi=$0.5, 1, 1.3, 1.8 and 2.4 for $h=800$). (b) Corresponding theoretical angular velocity signals. (c) Experimental (symbols) and theoretical (lines) amplitudes of the four first signal extrema versus $\xi$. (d) Experimental (symbols) and theoretical (lines) durations of the three first signal phases versus $\xi$. Error bars limits represent simulated results for $h=640$ and $h=960$. \label{fig4}}
\end{figure}

In Fig.~\ref{fig4}(a) we plot the experimental angular velocity signals for different excitation amplitudes at bead 1 ($z=0$ so by definition for $\xi=0$) and at bead 18 ($z=18d=0.323$ m), i.e. for different values of $\xi$. For the smallest excitation amplitude, the characteristic angular velocity is $\theta_0\simeq 0.17$ rad/s and the experimentally found speed of sound is $c=20\ $m/s. As observed and predicted for waves in resonance experiments \cite{nazarov,nazarov2,johnson,guyerprl99,johnsonsutin}, both effects of nonlinear softening and nonlinear attenuation are manifested here for the multi-phase pulses through stretching in time and normalized amplitude decrease with increasing excitation level, respectively, Fig.~\ref{fig4}(c,d). Additionally, it is striking to observe that different parts of the signal are not attenuated in the same way. For instance, the third extremum is attenuated faster than the second phase, being initially smaller in amplitude (see Fig.~\ref{fig1} (c)), which is counter-intuitive. This effect is explained by the model in Eqs.~(\ref{eq4}-\ref{eq6}) where the evolutions of signal phases amplitudes are coupled with the phase durations.
For $h=800\pm 160$, all these observations are quantitatively captured by the modeled signal distortion from Eqs.~(\ref{eq4})-(\ref{eq6}).
This value of parameter of hysteretic nonlinearity is different from the one evaluated above for the one contact - one bead system resonance $h\simeq 258$, although of the same order of magnitude. This deviation could be attributed to the differences in the setup (chain of 2 beads versus 70 beads) through the role of the weight, but also to the possibly different contact wears and their expected important influence on $h$ through the friction coefficient $\mu$.

{\it Conclusions.} Existence and propagation of pure torsional waves in a granular magnetic chain have been reported. The torsional waves show dispersive properties associated to the periodicity of the medium. Through a single contact nonlinear resonance experiment, we verified a nonlinear quadratic hysteretic relationship for the moment-angle (equivalent to stress-strain). Following this observation, we developed a set of equations based on the Preisach-Mayergoyz model of hysteresis to describe the torsional pulse propagation. The model's results compare very well with experimental distorted pulses observed for increasing excitation amplitude in a 70 bead long chain. The reported nonlinear transformation of pulse profile in a medium with hysteretic quadratic nonlinearity, essentially extends the historically observed distortion by quadratic nonlinearity \cite{mendousse} and the more recently observed distortion by cubic nonlinearity \cite{catheline}.

In particular, we found a giant nonlinear attenuation (or shift in time) for each phase of the pulse, which depends on the characteristics (amplitude, duration...) of the particular phase of the signal and sometimes on the relative amplitudes of the previous phases. This signifies a short term wave-memory effect. The observed giant nonlinear attenuation effects and the short term memory could become key components of nonlinear elastic wave control devices and logic elements \cite{boechler,Li,yang}.

The presented configuration could be used as a model medium to study fundamental nonlinear wave processes, e.g. frequency mixing, interaction of counter-propagating waves, self-modulation instability, still unexplored for hysteretic nonlinearity. At the same time, it could also be used to study novel nonlinear processes which do not exist with classical nonlinearities, like the nonlinear memory and the influence of the preparation of the initial hysteron distribution on the nonlinear pulse propagation.


\begin{thebibliography}{99}
\bibitem{note}Note here that due to the proportionality between the angular particle velocity $\bar{\theta}_v$ and the angular strain, the profiles of these quantities are the same and their evolution is described by Eq.~(\ref{eq3}).
\end{thebibliography}
\end{document}